%Paper: hep-ph/9308254
%From: logan@mich1.physics.lsa.umich.edu
%Date: Mon, 9 Aug 1993 16:02:38 -0400
%Date (revised): Wed, 11 Aug 1993 11:53:37 -0400
%Date (revised): Fri, 8 Oct 1993 11:26:06 -0400

%%%%%%%%%%%%%%%%%%%%%%%%%%%%%%%%%%%%%%%%%%%%%%%%%%%%%%%%%%%%%%%%%%%%
\input phyzzm

%%%%%%%%%%%%%%%%%%%%%%%%%%%%%%%%%%%%%%%%%%%%%%%%%%%%%%%%%%%%%%%%%%%%
%\input /marty/Tex/journals

\def\npb{{Nucl.\ Phys.\ }{\bf B}}

\def\plb{Phys.\ Lett.\ }

\def\prd{{Phys.\ Rev.\ }{\bf D}}

%%%%%%%%%%%%%%%%%%%%%%%%%%%%%%%%%%%%%%%%%%%%%%%%%%%%%%%%%%%%%%%%%%%%

\def\fourphoton{$\Pi_{\nu\rho\lambda\sigma}$}
\nopagenumbers
%\null
%\vskip-0.6in
%\nopubblock
\theory
\date{August 8, 1993}
\pubtype{hep-ph/9308254}  % a null string in the upper right
\pubnum{18} % UM-PHYSICS-86-747 is the result
\titlepage
%   \TITLEREF\onlyref{Present whereabouts unknown}
%\vskip1.5in
\title{\caps{On the Hadronic Contribution to Light-by-light Scattering in
$g_\mu-2.$}}
\vskip.75in
\author{Martin B. Einhorn}
\address{Randall Laboratory, University of Michigan, Ann Arbor, MI 48109-1120}
\abstract
We comment on the theoretical uncertainties involved in estimating the
hadronic effects on the light-by-light scattering contribution to the
anomalous magnetic moment of the muon, especially based on the analysis and
results of \REF\kno{T. Kinoshita, B. Ni\v zi\'c, and Y. Okamoto, \prd31, 2108
(1985).}Ref.~\kno.   From the point of view of an effective field theory and
chiral perturbation theory, we suggest that the charged pion contribution may
be better determined than has been appreciated.  However, the neutral pion
contribution needs greater theoretical insight before its magnitude can be
reliably estimated.

\endpage
\footline{\hss\tenrm\folio\hss}
\pagenumber=1
%\doublespace
The construction of an extremely high-precision experiment to determine the
anomalous magnetic moment of the muon, $a_\mu\equiv {(g_\mu-2)/2}$, is
underway at Brookhaven National Laboratory (BNL).\REF\gminustwo{F.J.M. Farley
and E. Picasso, ``The Muon $g-2$ Experiments" in {\sl Quantum Electrodynamics}
ed. T. Kinoshita, Singapore: World Scientific, 1990.}
\foot{For a summary and review of earlier measurements, see, e.g.,
Ref.~\gminustwo.} The anticipated design sensitivity, $\Delta a_\mu=4\times
10^{-10},$ will, if achieved, be about 20 times better than the results from
CERN,\refmark\gminustwo\ corresponding to a magnitude 5 times smaller than the
Standard Model weak corrections.  If the theory of known contributions is
sufficiently good, such a measurement would either constrain or reveal physics
beyond the Standard Model.\REFS\tkwm{T. Kinoshita and W.J. Marciano, ``Theory
of the Muon Anomalous Magnetic Moment" in {\sl Quantum Electrodynamics}, {\it
op.cit.}}\REFSCON\kinrev{T. Kinoshita,  Talk at 10th International Symposium
on ``High Energy Spin Physics," Nagoya, Nov.\ 9-14, 1992, (CLNS
93/1186.)}\REFSCON\aew{C. Arzt, M.B. Einhorn, and J.  Wudka, Michigan
UM-TH-92-17 (Phys.\ Rev.\ D, to be published.)}\refsend  To draw such an
inference, or even to check the weak
radiative correction, requires a highly accurate determination of hadronic
contributions.  The $O(\alpha^2)$ correction to $a_\mu$ coming from the
$O(\alpha)$ contribution to hadronic vacuum polarization to the photon
propagator, has received a great deal of attention\refmark\tkwm\ since it is
approximately 35 times larger than the weak correction.  The systematic error
on this contribution can be reduced to the level of accuracy of the BNL
experiment by a more precise low-energy measurement of
$R\equiv\sigma(hadron)/\sigma(\mu^-\mu^+),$ at VEPP-2M,\refmark\kinrev\
currently operating, or, in the future, at DA$\Phi$NE.\Ref\daphne{{\sl The
Da$\Phi$ne Physics Handbook} ed. L. Maiani et.al., Frascati, INFN, 1992.}

%{\it What about the $O(\alpha^2)$ contributions to vacuum polarization?
%There appear to be contributions omitted corresponding to associated
%bremsstrahlung and virtual electromagnetic corrections.}

There are other hadronic contributions in $O(\alpha^3)$ coming from the
$O(\alpha^2)$ contribution to the four-photon vacuum polarization tensor
\fourphoton, the so-called contribution from light-by-light
scattering.  At first sight, since hadronic interactions are strong, this
would appear to be very difficult to determine accurately, but Kinoshita
et.al.\refmark\kno\ have estimated this contribution to be $49\times 10^{-11}$
with an quoted accuracy of about $10\%.$  If correct, this contribution is a
bit larger than the anticipated accuracy of the experiment, but four times
smaller than the Standard Model weak correction.  The method of calculation
employed elementary pion contributions to the hadronic amplitude, together
with vector meson dominance (VMD) of the photon couplings to the pion.  They
noted that their results happened to agree with a one-loop calculation
involving elementary quarks with constituent quark masses, but they did not
place great store by this second method because it is sensitive to the choice
of the quark mass and would not appear to be a correct physical picture for
momenta on the order of $m_\mu$ or so.

Barbieri and Remiddi\Ref\rber{R. Barbieri and E. Remiddi, {\sl The DA$\Phi$NE
Physics Handbook,} {\it op.cit.} vol.\ II, p. 301.} have recently raised
doubts about the degree of certainty to be associated with the magnitude of
the hadronic contribution in light-by-light scattering.  As this is important
for the interpretation of the experiment and crucial for drawing inferences
about potential new physics, it is the purpose of this note to stimulate
further discussion of these important issues.

The criticism in Ref.~\rber\ is two-fold: (1)~It is argued that the estimation
in Ref.~\kno\ by a quark loop is untrustworthy since it is so sensitive to the
value chosen for the (constituent) quark mass.  In this respect, they are
actually in agreement with Ref.~\kno, who did {\bf not} base their errors on
the agreement with the estimation via a quark loop.  So this should not be a
point of dispute.  (2)~They doubt the accuracy claimed in Ref.~\kno, because
it results ``from the cancellation of different (gauge invariant)
contributions, each of which is larger than the electroweak correction itself,
...."\refmark\rber\  In this regard, we believe that they are mistaken in
suspecting that the approximation is unstable.  In fact, the separate
contributions, labelled A, B, and C in Ref.~\kno, while gauge-invariant with
respect to the external photon, are not gauge-invariant contributions to
\fourphoton; in particular, Bose symmetry among the four photons, while true
of the sum, is not respected by the three separate sets of diagrams.  Although
each set, A, B, and C, yields a finite contribution to $a_\mu,$ it commonly
happens in the calculation of radiative corrections that there are large
cancellations among classes of diagrams that are not separately gauge
invariant, and such an occurrence in and of itself is not necessarily cause to
distrust the final result.

Let us consider, {\it ab initio}\/ how one may approximate the hadronic
contributions to \fourphoton.  The relevant momentum scale for the external
photon momenta attached to \fourphoton\ is set by the muon mass, an energy
scale small compared to typical hadronic scales, with the exception of the
pion.  Therefore, it should be a good approximation to describe this by an
effective field theory involving pions and photons only.  At low energies, the
pions may be regarded as the Goldstone bosons associated with chiral symmetry
breaking, whose self-interactions are described by the $O(3)$ non-linear sigma
model, gauged with respect to the $U_1$ of electromagnetism.  As a result, the
interactions among pions at low-energy are actually weak, the scale of the
chiral perturbation expansion being set by the mass $m_\rho$ of the $\rho$
meson or thereabouts, well above the mass of the muon or pion.  As a first
approximation, then, one may neglect the self-interactions of pions entirely.
For the charged pion, this is precisely what was done in Ref.~\kno, but one
must understand that this is not a {\it model} for
\fourphoton, but the first term of a systematic expansion.  The result for
this contribution in Ref.~\kno\ was given in Eq.~(3.20).  To illustrate that
there should be no mystery here, a back-of-the-envelope estimate for this
contribution to $a_\mu$ is as follows:  all such contributions are
proportional to $m_\mu^2$: one power comes from the definition of $a_\mu$;
the other comes from the observation that the photon couplings are all chiral
conserving whereas the anomalous moment requires a chiral-flip, so a mass
insertion is required.  This contribution will then be proportional to
$(m_\mu/m_\pi)^{-2}.$  There are 6 factors of the coupling $e$ and a factor of
$1/16\pi^2$ for each loop.  While each graph contributes a logarithmic
subdivergence to \fourphoton, we know from gauge invariance that this
divergence is cancelled in the sum.  So we estimate
$$(m_\mu/m_\pi)^2(\alpha/4\pi)^3\approx .01(\alpha/\pi)^3.\eqn\eq$$ The actual
result of $-.04(\alpha/\pi)^3,$ given in Eq.~(3.20) of Ref.~\kno, supports the
view that, even though there are 21 graphs to be added,\foot{Actually there
are only 8 different integrals; see Fig.~5 of Ref.~\kno.} the cancellation
among divergent pieces also holds for the finite remainder.  Thus, the final
result is no bigger than our estimate, reinforcing our belief that the
cancellation between subsets of graphs is simply a gauge cancellation.  We do
not have an argument for the sign of the contribution.

Leaving aside the matter of the chiral anomaly for a moment, what are the
corrections from interactions among the pions?  The pion interactions do not
become strong until one reaches to vicinity of the $\rho$-meson, so one would
expect such contributions to $g_\mu-2$ to be suppressed compared to the free
pion contribution by a factor of approximately $m_\pi^2/m_\rho^2\approx 4\%.$
In Ref.~\kno, some of these corrections are introduced by VMD for the photon
propagators, and it was found that they make corrections of more than 30\% to
various groups of diagrams contributing to $a_\mu$ and reduce the final answer
for the charged pion contribution by a factor of more than 3.  This casts a
seemingly reliable approximation scheme in doubt.  The VMD approximation used
in Ref.~\kno\ simply multiplies \fourphoton\ by a common factor.  Therefore,
\fourphoton\ remains gauge invariant,\foot{I thank T. Kinoshita for emphasizing
this point.} even th
ough this VMD prescription does not respect the Ward identities for the
couplings of photons to charged pions when off-mass-shell.\REF\lpdr{G. Ecker
et.al., Phys. Lett. B223, 425 (1989).}\foot{In fact, chiral perturbation theory
has been extended to inclu
de the $\rho,$\refmark\lpdr but it is not clear how that formalism would be
helpful here.}    Since \fourphoton\ produces a convergent integral for the
free pion loop, it is hard to understand how the VMD approximation, modifying
the integrand on scales l
arge compared to the pion mass, can change the final answer by a factor of
three.  This is a puzzle which suggests that the numerical integration be
confirmed.   It would add great confidence to the error estimates if the VMD
corrections could be recalcul
ated and found to be of order $m_\pi^2/m_\rho^2$ as expected.  Of course, the
integrand is not positive definite, so if the  low-momentum region contributed
little to the final answer, then modification of the high-momentum regime
conceivably could produc
e a large percentage change in the final result.  However, in that case, it may
be important to have a gauge-invariant VMD model.  For the time being, we tend
to trust the free pion result more.

Another approach to estimating corrections would be to determine the effects
of the higher order terms in the chiral perturbation expansion.  That would be
a complicated calculation that we have not attempted.  However, we may
anticipate one feature of these corrections, viz., because they involve
higher-derivative interactions, they will lead to {\it divergent}
contributions to $a_\mu.$  That simply signifies that, when the muon is
included in the effective field theory, one must include a ``direct"
contribution to $g_\mu-2$ for which such a divergence provides a
renormalization.\foot{This sort of contribution to $a_\mu$ and the
corresponding effective field theory is described in detail in Ref.~\aew.  It
occurs also for the $\pi^0$ contribution; see below.} That is to say, the
effective field theory contains a term of the form $${
\alpha_d\over{\Lambda^2}}(\bar\psi \sigma^{\mu\nu} \psi)
F_{\mu\nu},\eqn\direct$$ where $\alpha_d$ represents some effective coupling
constant and $\Lambda$ represents the scale at which this effective field
theory breaks down.  In the present context, we would expect $\Lambda$ to be
on the order of the masses of the vector mesons $m_\rho$ and $m_\omega.$  It
has been tacitly {\it assumed} in Ref.~\kno\ that the direct contribution is
negligibly small.\foot{We shall revisit this assumption below in connection
with the neutral pion contribution.} Strictly speaking, one requires a
renormalizable description of strong interactions in order to be able to
assume that the bare $\alpha_d$ is zero.  QCD is, of course, such a theory
but, unfortunately, we cannot reliably calculate its behavior in the
low-energy regime relevant here.  Even if we could, one must expect the
physics omitted from any approximate description, renormalizable or not, to
produce a direct interaction of this sort.   For example, the weak correction
itself may be regarded as a direct contribution at scales below
$M_W$.\Ref\witten{E. Witten, \npb122, 109 (1977).}  One might explore various
renormalizable models for chiral-symmetry-breaking that are somewhat simpler
than QCD;  for example, one might choose the {\it linear} sigma model, gauged
with respect to electromagnetism.  This effectively provides a cutoff at the
mass of the $\sigma,$ and the sensitivity of the result to $\sigma$ mass (or
the strength of the pion self-coupling) may provide an indication of the size
of the uncertainty in the corrections; however, extracting the contribution to
$a_\mu$ from a two-loop contribution to \fourphoton\ is a bit daunting.

In sum, we suggest that the result given for the non-interacting pion loop,
Eq.~(3.20) of Ref.~\kno, may be a reliable estimate of the charged pion
contribution to an accuracy of about 4\%.  The VMD model is not
gauge-invariant, although it does leave \fourphoton\ gauge invariant.   As a
result, we would be inclined to trust the non-interacting result more than the
somewhat smaller one favored in Ref
.~\kno\ resulting from their implementation of VMD.  We now turn to consider
the other contribution at the pion mass scale, the contribution of the $\pi^0.$

The effective field theory must be amended to include the effect of the chiral
anomaly by adding\Ref\slawab{S.L. Adler and W.A. Bardeen, Phys.\ Rev.\ {\bf
182} 1517 (1969).} $$G{\alpha\over{8\pi f_\pi}
}\pi^0\widetilde{F}_{\mu\nu}F^{\mu\nu}\eqn\anomaly$$ in the linear
representation or the corresponding expression for the nonlinear realization
in chiral perturbation theory.\Ref\wzw{J. Wess and B. Zumino, \plb37B, 95
(1971), E. Witten, \npb223 422 (1983).}  (Here, the pion decay constant
$f_\pi\approx~$93~MeV.) Eq.~\anomaly\ describes the coupling of the $\pi^0$ to
two photons at low energy, and this local vertex was used as the first
approximation in Ref.~\kno\ to the contribution of the $\pi^0$ to \fourphoton.
The coupling constant $G$ is predicted to be 1 in QCD in the ``chiral limit"
in which the pion is massless, and the experimental value agrees extremely
well with this, to about 0.25\%.  Power counting shows that this local
approximation to the interaction leads to a (logarithmically) divergent result
for $a_\mu.$  This point-like interaction will be damped on scales associated
with pion substructure, so, at scales on the order of $m_\rho,$ this local
approximation to the coupling $G$ must be modified.  This is qualitatively
equivalent to introducing damping factors via VMD as in Ref.~\kno, with a
result that is given in their Eq.~(3.27).  In this case, their approximation
to the vertex is gauge-invariant, and, therefore, this can be expected to
provide a believable estimate of the size of this correction.  The result
should not be very different from simply introducing a cutoff on the
divergence at the scale $m_\rho,$ and, since the divergence is only
logarithmic, the result will not be very sensitive to the precise value
chosen.  Again, a back-of-the-envelope calculation gives for any one diagram
$$a_\mu\sim m_\mu^2 ({1\over{16\pi^2}})^2 {e^2}({{\alpha}\over{\pi f_\pi} })^2
\ln(m_\rho^2/m_\pi^2)= {1\over4}({ {m_\mu}\over{4\pi f_\pi} })^2
({\alpha\over\pi})^3\ln(m_\rho^2/m_\pi^2) \approx .006(\alpha/\pi)^3.$$  There
are two equal pairs of two diagrams each,\foot{The two diagrams of Fig.~5 of
Ref.~\kno\ must be added to two having the muon line reversed.} so we ought to
multiply this by at least a factor of two;  that is still about a factor of
4 less than the result of $.05(\alpha/\pi)^3$ given in Eq.~(3.27) of
Ref.~\kno, using VMD to cut off the divergence.

It is actually quite surprising that the magnitude of the $\pi^0$ contribution
is as large as the charged pion contribution.\foot{According to Ref.~\kno, the
free pion result is about equal but of opposite sign.} Even though it is
obviously the same order in $\alpha,$ it is in effect one-loop order higher.
That is, the anomalous coupling $\alpha/{\pi f_\pi}$ itself is a one-loop
contribution, and it enters squared in a two-loop graph.  Thus, this is in
effect a four-loop contribution, to be compared with the three-loop
contributions involving the charged pion.  So, {\it a priori,} one would
anticipate this contribution being suppressed by $1/16\pi^2.$   Even though
there are many more diagrams in the charged pion case, the cancellations among
diagrams as required by gauge invariance arranges for the final result to be
no bigger than our estimate.  Thus, whereas the charged pion contribution is
somewhat smaller than one might have guessed given the number of diagrams, the
$\pi^0$ contribution is somewhat larger.  This enhancement may be attributed
to a host of small factors: the anomaly is twice as large as we would have
guessed and it enters squared; the log enhancement is about a factor of 3; a
factor of 2 from $(m_\pi/f_\pi)^2;$ finally the various diagrams may add
constructively in this case and destructively in the charged pion case.  Thus,
remarkably, the additional factor of $1/16\pi^2$ may be largely compensated,
but it remains rather surprising and would be worth double-checking.

Combining the neutral pion result with the charged pion loop contribution
(Eq.~(3.20) rather than Eq.~(3.24) of Ref.~\kno,) one obtains the result
$$a_\mu({\rm had2})=0.014\Bigl[{\alpha\over\pi}\Bigr]^3,\eqn\eq$$
coincidentally about 3 times {\it smaller} than that given there.  This
depends upon the significant cancellation between the charged and neutral pion
contributions.

If this were the whole story, the situation would be quite satisfactory.
However, the fact that the local coupling Eq.~\anomaly\ results in a
ultraviolet divergence implies that there must also be a ``direct"
contribution to $a_\mu$ coming from physics above the cutoff ($m_\rho$) whose
natural size cannot be much smaller than the radiative correction due to the
$\pi^0.$  There are several ways to see this: If we regard $\alpha_d$ as a
bare coupling, then the divergence of the $\pi^0$ contribution renormalizes
it.  Alternatively but equivalently, we may infer from the logarithmic
divergence a contribution to the $\beta$ function for the renormalized
coupling $\alpha_d(M),$ where $M$ is the renormalization scale.  It would be
{\it unnatural,} in the technical sense of the word,\Ref\wein{S. Weinberg,
Physica {\bf 96A} 327 (1979).} for it to be small compared to the size implied
by varying $M$ a bit over the range of interest.  Since $\ln
(m_\rho/m_\pi)\approx 1.6,$ we would expect the direct term to be not much
smaller than the $\pi^0$ contribution calculated in Ref.~\kno.  {\it This
suggests that important physics has been omitted from this method of
calculation whose magnitude may well be comparable to the contribution
calculated.}

Of course, we expect that the $\pi^0\gamma\gamma$ vertex, even for the simple
triangle diagram, will be damped whenever one of the external momenta becomes
large compared to the typical scale of strong interactions.  On the other
hand, the Adler-Bardeen theorem\refmark\slawab\ guarantees that, for zero
external momenta, the strong interactions do not in fact modify the anomaly
associated with the underlying fundamental quark-loop diagram.  The
theoretical challenge is to figure out how to move away from the local
approximation in a manner that is sufficiently well-controlled to provide an
estimate of the error on the calculated contribution to $g_\mu-2$.

The implications of this uncertainty associated with the $\pi^0$ contribution
are severe: If the data turn out not to agree with the theoretically
predicted value and the experimental errors are sufficiently small, then,
rather than necessarily a signature of new physics, the deviations could just
as well be ascribed to an erroneous assumption about the magnitude of the
hadronic contribution from light-by-light scattering.  In this respect, we
agree with Ref.~\rber\ that more work needs to be done.

\ack
I have enjoyed discussions on this subject with T. Kinoshita and D.R.T. Jones.
This work has been supported in part by the U.S. Department of Energy.

\vfill\eject
\refout
%\figout
\end